\documentclass[preprint,12pt,times]{elsarticle}
\usepackage{graphicx}
\usepackage[dvipsnames,rgb,dvips]{xcolor}
\usepackage{amsmath,amssymb}
\usepackage{bm}

\usepackage{url}
\definecolor{blue}{rgb}{0,0,1}

\usepackage[title]{appendix}
\newcommand{\ve}[1]{\ensuremath{\mbox{\boldmath$#1$}}}
\newcommand{\ma}[1]{\ensuremath{\mathbb{#1}}}

\newcommand{\st}{\ensuremath{\mbox{St}}}
\newcommand{\steff}{\ensuremath{\widetilde \st}}
\newcommand{\pe}{\ensuremath{\mbox{Pe}}}
\newcommand{\re}{\ensuremath{\mbox{Re}}}

\newcommand\transpose{^{\mathrm T}}

\newcommand{\spma}[1]{#1\ve n - \ve n(\ve n\transpose #1\ve n)}
\newcommand{\venn}{\ve n \ve n\transpose}
\newcommand{\maid}{\ma I}

\newcommand{\evalat}[1]{\,\bigg|_{#1}}
\newcommand{\rd}{\ensuremath{\mathrm{d}}}
\newcommand\nn{\nonumber}
\newcommand{\eqnlab}[1]{\label{eqn:#1}}

\newcommand{\Eqnref}[1]{Eq.~(\ref{eqn:#1})}

\newcommand{\Ylm}[2]{|#1,#2\rangle}
\newcommand{\cross}{\wedge}

\journal{Physica D}
\begin{document}

\begin{frontmatter}
\title{Orientational dynamics of weakly inertial axisymmetric particles in steady viscous flows}

\author[ugo]{J. Einarsson}
\author[cae]{J. R. Angilella}
\author[ugo]{B. Mehlig}
\address[ugo]{Department of Physics, University of Gothenburg, SE-41296 Gothenburg, Sweden}
\address[cae]{Department of Mathematics and Mechanics, LUSAC-ESIX, University of Caen, France}

\begin{abstract}

The orientational dynamics of weakly inertial axisymmetric particles in a steady flow is investigated.
 We derive an asymptotic equation of motion for the unit
axial vector along the particle symmetry axis, valid for small Stokes number $\st$, and for any axisymmetric particle in any steady linear viscous flow. This reduced dynamics is analysed in two ways, both pertain to the case of a simple shear flow. In this case inertia induces a coupling between precession and nutation. This coupling affects the dynamics of the particle, breaks the degeneracy of the Jeffery orbits, and creates two limiting periodic orbits. We calculate the leading-order Floquet exponents of the limiting periodic orbits and show analytically that prolate objects tend to a tumbling orbit, while oblate objects tend to a log-rolling orbit, in agreement with previous analytical and numerical results. Second, we analyse the role of the limiting orbits when rotational noise is present. We formulate the Fokker-Planck equation describing the orientational distribution of an axisymmetric particle, valid for small $\st$ and general P\'eclet number $\pe$. Numerical solutions of the Fokker-Planck equation, obtained by means of expansion in spherical harmonics, show that stationary orientational distributions are close to the inertia-free case when $\pe\st\ll1$, whereas they are determined by inertial effects, though small, when $\pe \gg 1/\st\gg1$.
\end{abstract}

\begin{keyword}
shear flow, Jeffery orbits, particle inertia, Floquet analysis, Fokker-Planck equation
\end{keyword}
\end{frontmatter}

\section{Introduction}

Suspensions of rigid particles are abundant in both nature and technology, and are consequently studied in many disciplines of science. Examples are 
ash and ice crystals in the atmosphere \cite{Rolf2012}, plankton in the ocean \cite{Stommel1949, Gua12}, or fibres in paper pulp \cite{Lun11}. In many situations the concentration of the suspended particles in the carrying fluid is small, so that particles
 do not affect the underlying fluid flow. In this case, considering the fluid flow field as given, 
 a question of fundamental interest is to determine 
 the resulting trajectory and orientation of a particle.  The orientational dynamics of axisymmetric particles is often described by the Jeffery equation of motion \cite{Jeff22}, 
the rotational analogue to simple center-of-mass advection along streamlines. 
This equation of motion is valid in the limit of vanishing Stokes number $\st$ and particle Reynolds number $\re$ (both numbers are defined below), representing particle and fluid inertia respectively. The Jeffery equation of motion is widely used to predict mechanical \cite{Bre74,Lun11,pet99} as well as optical properties \cite{Gauthier1998,wil09,wil11,Bez10} of dilute suspensions of inertia-free rod-like or disk-like particles. It is also used to describe the orientational motion of single particles \cite{Gua12,ein12}, and the alignment and tumbling of axisymmetric particles in turbulence \cite{Pum11,Par12}.

In the case of simple shear flow an inertia-free particle rotates in one of infinitely many possible Jeffery orbits. In other words, the solution of the Jeffery equation is degenerate, forming a one-parameter family of closed orbits. Which orbit is followed depends upon the initial condition.
By taking into account the inertia of the particle in numerical simulations, it was observed that attracting orbits exist \cite{Lundell2010,Lundell2011}, where particles tend to align their small axis towards the vorticity axis. In the special case of nearly spherical particles the existence of attracting orbits has been demonstrated analytically \cite{Sub06}. The significance of this inertial effect is that it breaks the degeneracy of the Jeffery orbits.

If, in addition, particles are subject to a random force, the resulting orientational distribution is expected to be determined by a balance between the dispersing random forces and the aligning inertial forces. 

The goal of the present study is to investigate the orientational dynamics of axisymmetric particles with small inertia, to describe and quantify the attracting orbits, and to analyse their effect on the orientational distribution when noise is present.

To accomplish this goal we first derive an equation of motion
 for the orientation of an axisymmetric particle in steady flows in the absence of noise, valid for small values of the Stokes number $\st$.
We define the Stokes number by $\st = ms/b\mu$, where $m$ is the particle mass, $s$ is a typical flow-gradient rate, $b$ is the minor particle length and $\mu$ the dynamic viscosity of the fluid (see Sec.~2 for details). 
We make use of the fact that $\st$ is small, so that the angular velocity of the particle is a small perturbation of the angular velocity in the absence of particle inertia.
Our results, valid for any aspect ratio, generalise a result for nearly spherical spheroids reported in Ref.~\cite{Sub06}. In Ref.~\cite{Sub06} fluid inertia, as measured by the particle Reynolds number, is also considered.
The particle Reynolds number is defined by $\re = (\kappa\rho_f/\rho_p)\st$, where $\rho_f$ and $\rho_p$ denote fluid and particle density. The \lq shape factor\rq{} $\kappa$ is 
$\kappa=1$ for a disk-shaped particle, and $\kappa=\lambda$ for a rod-like particle ($\lambda$ is the aspect ratio of the particle). Finite-$\re$ corrections are important for neutrally buoyant particles [$\rho_p/\rho_f=\mathcal O(1)$], because the fluid inertia induces $\mathcal O(\re)$-corrections in the hydrodynamic torque experienced by the particle \cite{Sub06,lov93}. 
The results obtained here are valid for heavy particles such that $\rho_p \gg \kappa\rho_f$.

The new equation of motion allows us to derive a Fokker-Planck equation determining the combined effects of particle inertia and noise
upon the orientational distribution of the particles. This equation describes the evolution of an ensemble of weakly inertial particles in a steady flow, subject to random Brownian rotations. It is solved in the case of a simple shear flow by means of expansion in spherical harmonics. We find that the stationary orientational distributions differ significantly from the inertia-free case.

   The rest of this paper is organised as follows. Asymptotic equations of motion valid for small $\st$ in the absence of noise
are obtained  in Sec.~2. Then, in Sec.~3, we investigate the case of a simple shear flow. We evaluate the convergence towards one of two possible periodic orbits by calculating the corresponding Floquet exponents. Finally, in Sec.~4, we formulate the Fokker-Planck equation describing the combined effect of noise and weak particle inertia. This equation is then solved numerically and the stationary distribution is discussed. Sec.~\ref{sec:conc} summarises our conclusions.

\section{Equation of motion}
The aim of this Section is to derive an equation of motion for the orientation of a small axisymmetric particle, valid to first order in $\st$. The orientation is described by a unit vector $\ve n$ pointing in the direction of the symmetry axis of the particle. In the following we derive an equation of the form $\dot{\ve n}=\ve F_0(\ve n) + \st \ve F_1(\ve n) + \mathcal O (\st ^2)$, where the dot denotes the time derivative, and the functions $F_i$ depend on the flow gradient $\ma A = \ve \nabla \ve u$. We decompose $\ma A$ into its symmetric and antisymmetric parts
\begin{align*}
	&\ma O = \frac{1}{2}(\ma A - \ma A\transpose),\quad
	\ma S = \frac{1}{2}(\ma A + \ma A\transpose),\quad
	\ma A = \ma O + \ma S.
\end{align*}
The matrix $\ma O$ is related to the vorticity $\nabla \cross \ve u$ of the flow: $\ma O \ve x = \ve \Omega \cross \ve x$, where $\ve \Omega = \nabla \cross \ve u/2$.

We start with Newton's equation of motion for the orientational degrees of freedom for a rigid body. Let $\ve \omega$ denote the angular velocity of the particle, then
\begin{align}
	\dot{\ve \omega} &= \ma M^{-1}\left(-\dot{\ma M}\ve \omega + \ve T\right) \nn\\
	\dot{\ve n} &= \ve \omega \cross \ve n\eqnlab{fulleqns}
\end{align}
where $\ma M$ is the moment of inertia, and $\ve T$ the torque applied to the body. The hydrodynamic torque on a particle depends on the flow configuration at the particle position. Therefore, when studying inhomogenous flows such as turbulent channel flows, \Eqnref{fulleqns} must be completed with equations for the particle center-of-mass motion \cite{Mar10}. Here we consider homogeneous flows where the center-of-mass motion may be neglected. With the hydrodynamic torque calculated by Jeffery \cite{Jeff22}, \Eqnref{fulleqns} describes the rotation of both axisymmetric \cite{Lundell2010} and asymmetric \cite{Lundell2011} particles. In the following we consider the case of axisymmetric particles.

An axisymmetric object has two principal moments of inertia, around and transverse to the axis of symmetry. The moment of inertia tensor is therefore on the form
\begin{align}
	\ma M &= \widetilde{X}^I\venn + \widetilde{Y}^I (\maid - \venn),\eqnlab{momentofinertia}
\end{align}
where $\maid$ is the identity matrix, and $\widetilde{X}^I, \widetilde{Y}^I$ are constants depending on the shape of the particle. Here and in the following our notation is similar to that used in the book by Kim and Karrila \cite{Kim91}. The difference, and the reason we keep the tildes, is that here the quantities have dimensions. For instance a spheroid with lengths $a$ and $b$ and mass $m$ has $\widetilde{X}^I=2mb^2/5$ and $\widetilde{Y}^I=m(a^2+b^2)/5$. Further, the hydrodynamic torque depends linearly on the flow gradient via resistance tensors which are determined solely by the shape of the particle \cite{Bre74,Jeff22, Kim91}
\begin{align}
	\ve T &= \left[\widetilde{X}^C \venn + \widetilde{Y}^C(\maid - \venn)\right](\ve \Omega-\ve \omega)+\widetilde{Y}^H \ve n \cross \ma S \ve n.\eqnlab{torque}
\end{align}
Here $\widetilde{X}^C, \widetilde{Y}^C$ and $\widetilde{Y}^H$ are shape constants. The relation to the dimensionless shape functions given in \cite{Kim91} is, for instance, $\widetilde{X}^C=8\pi\mu a^3 X^C$. The shape functions for spheroids are given on p.\,64 (prolate case) and p.\,68 (oblate case) of Ref.~\cite{Kim91}. Inserting the explicit forms \Eqnref{momentofinertia} and \Eqnref{torque} into \Eqnref{fulleqns} yields
\begin{align}
	\dot{\ve \omega} &= \left[\frac{\widetilde{X}^C}{\widetilde{X}^I}\venn + \frac{\widetilde{Y}^C}{\widetilde{Y}^I}(\maid-\venn)\right](\ve \Omega - \ve \omega) \nn\\
	&\quad+ \frac{\widetilde{Y}^H}{\widetilde{Y}^I}\ve n\cross\ma S \ve n -\frac{\widetilde{X}^I-\widetilde{Y}^I}{\widetilde{Y}^I}(\ve \omega \cross \ve n)\ve n\transpose\ve \omega.\eqnlab{dimomega}
\end{align}
An axisymmetric body has two different lengths $a > b$, and we choose to base the Stokes number on the lesser of the two, such that $\st=ms/b\mu$. Here $m$ is the mass of the particle, $s$ is a typical shear rate and $\mu$ is the dynamic viscosity of the surrounding fluid. We introduce dimensionless variables $t=t'/s$, $\ma A = s\ma A'$, $\ve \omega = s \ve \omega'$, and drop the primes.

\begin{figure}
\includegraphics{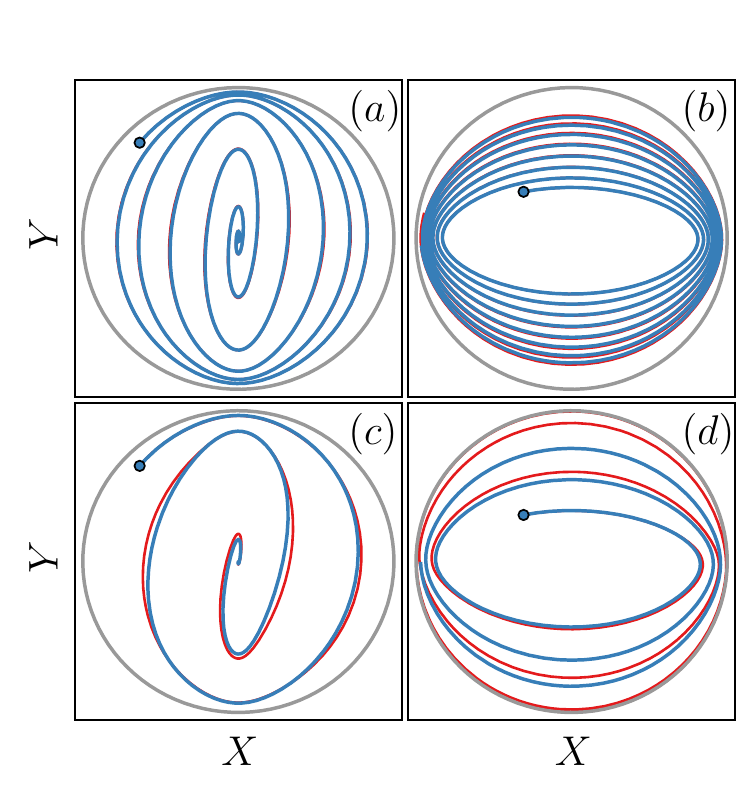}
\caption{\label{fig:eqcheck} Numerical solutions of \Eqnref{fulleqns} (red), and \Eqnref{nequation} (blue) for four different particles in a simple shear flow.
(a) $\lambda=1/5$, $\steff=0.1$; 
(b) $\lambda=5$, $\steff=0.25$; 
(c) $\lambda=1/5$, $\steff=0.25$; 
(d) $\lambda=5$, $\steff=0.7$.
Trajectories on the sphere are shown projected onto a disk with the area-preserving Lambert azimuthal stereographic projection $X=n_x\sqrt{2/(1+n_z)}$, $Y=n_y\sqrt{2/(1+n_z)}$ \cite{sny87}. The grey boundary marks the equator, the blue dot is the initial condition.}
\end{figure}

We now assume that for small $\st$ the angular velocity $\ve \omega$ is almost the same as in the unperturbed case:
\begin{align*}
	\ve \omega = \ve \omega_0 + \st \ve \omega_1 + \mathcal{O}(\st^2).
\end{align*}
Inserting into \Eqnref{dimomega} and comparing order by order in $\st$ we find
\begin{align}
	0 &= \left[\alpha^X \venn + \alpha^Y(\maid-\venn) \right](\ve \Omega - \ve \omega_0) + \alpha^S \ve n \cross \ma S \ve n, \eqnlab{order0}\\
	\dot{\ve \omega}_0 &= -\left[\alpha^X \venn + \alpha^Y(\maid-\venn) \right]\ve \omega_1 + \alpha^I (\ve \omega_0\cross\ve n)\ve n\transpose\ve \omega_0 \eqnlab{order1}.
\end{align}
Here the dimensionless coefficients $\alpha^i$, $i=X$, $Y$, $S$, $I$ are defined by
\begin{align*}
	&\frac{s}{\st}\alpha^X \equiv \frac{\widetilde{X}^C}{\widetilde{X}^I},
	\quad \frac{s}{\st}\alpha^Y \equiv \frac{\widetilde{Y}^C}{\widetilde{Y}^I}, 
	\quad \frac{s}{\st}\alpha^S \equiv \frac{\widetilde{Y}^H}{\widetilde{Y}^I},
	\quad \alpha^I \equiv \frac{\widetilde{Y}^I-\widetilde{X}^I}{\widetilde{Y}^I}.
\end{align*}
The lowest order, $\ve \omega_0$ is found from \Eqnref{order0} to be
\begin{align*}
	\ve \omega_0 &= \ve \Omega + \frac{\alpha^S}{\alpha^Y} \ve n \cross \ma S \ve n
\end{align*}
which we recognise as the Jeffery angular velocity \cite{Kim91}. In particular, for a spheroid of aspect ratio $\lambda=a/b$ the material constant $\alpha^S/\alpha^Y=(\lambda^2-1)/(\lambda^2+1)\equiv \Lambda$. To recover the well-known Jeffery equation for $\dot{\ve n}$, note that
\begin{align*}
	\dot{\ve n} &= \ve \omega_0 \cross \ve n + \mathcal O (\st)
	= \ma O \ve n + \frac{\alpha^S}{\alpha^Y}(\ma S \ve n - \ve n (\ve n\transpose \ma S \ve n))  + \mathcal O (\st).
\end{align*}
Solving for the next order $\ve \omega_1$ in \Eqnref{order1} we find
\begin{align}
	\ve \omega_1 &= \frac{\alpha^I}{\alpha^Y}(\ve \omega_0 \cross \ve n)\ve n\transpose \ve \omega_0 \nn \\ 
	&\quad- \left[\frac{\alpha^S}{\alpha^Y \alpha^X}\venn + \frac{\alpha^S}{\alpha^Y \alpha^Y}(\maid-\venn)\right]\left((\ve \omega_0 \cross \ve n)\cross\ma S \ve n \right) \nn\\
	&\quad- \frac{\alpha^S}{\alpha^Y \alpha^Y}\ve n \cross \ma S(\ve \omega_0 \cross \ve n).\eqnlab{w1general}
\end{align}
This expression for the correction to the angular velocity is valid for any axisymmetric body, specified by the parameters $\alpha^i$. For the case of a simple shear flow and nearly spherical spheroids \Eqnref{w1general} agrees with earlier results (Eq.~(3.12) in Ref.~\cite{Sub06} is equal to \Eqnref{w1general} to first order in $\Lambda$). In the case of spheroids, both prolate and oblate, we insert the appropriate shape functions from \cite{Kim91} and observe that
\begin{align*}
	&\alpha^I = \frac{\alpha^S}{\alpha^Y} = \frac{\lambda^2-1}{\lambda^2+1}\equiv \Lambda.
\end{align*}
We find our final result for the equation of motion for $\ve n$ to order $\mathcal O(\st)$ as
\begin{align}
	\dot{\ve n} &= \ma O \ve n + \Lambda(\spma{\ma S}) \nn\\
	&+\nn \frac{\st \Lambda}{\alpha^{Y}}\bigg[-\left( \spma{(\ma O\ma O+\ma S\ma O+\Lambda\ma S\ma S)} \right)\\\nn
	&\qquad\qquad-\Lambda\ve n\transpose \ve \Omega\left(\ve n \cross \ma S \ve n\right) \\\nn
	&\qquad\qquad+(\ve n\transpose\ma S \ve n)\ma O \ve n \\
	&\qquad\qquad+2 \Lambda (\ve n\transpose\ma S \ve n)(\spma{\ma S})\big] + \mathcal O(\st^2)\,. \eqnlab{nequation}
\end{align}
\Eqnref{nequation} is valid as an approximation to \Eqnref{fulleqns} when particles are weakly inertial.
The vector $\ve n$ is advected in an effective vector field $\ve F_0(\ve n) + \st\ve F_1(\ve n)$ that differs from Jeffery's equation by a term linear in $\st$. This approximation does not capture the formation of caustics that occur when particles with different orientational histories arrive at the same orientation but with different angular velocities \cite{gus13}. From \Eqnref{nequation} we see that the geometrical factor multiplying the inertial correction is $1/\alpha^Y$. Therefore, in the following, we use the effective Stokes number $\steff = \st/\alpha^Y$. Solutions to the full equation of motion \Eqnref{fulleqns} are compared to solutions of \Eqnref{nequation} for a simple shear flow and varying $\lambda$ and $\steff$. We observe that the approximation works well for the examples shown in Fig.~\ref{fig:eqcheck}, although the deviation in panel (d) is discernible.

\section{Simple shear flow: stability analysis}
The simple shear flow, $\ve u(\ve r, t)= s y \hat{\ve x}$, is interesting because it is the local linearisation of any flow with parallel streamlines, like pipe flows. It is also the flow in an ideal Couette rheometer used to measure the viscosity of fluids. The solutions of \Eqnref{nequation} in the simple shear at $\st=0$ are a one-parameter family of closed periodic orbits, the well-known Jeffery orbits \cite{Jeff22}. The effect of the $\mathcal O(\st)$ terms in \Eqnref{nequation} is a drift towards a stable periodic orbit. In this section we will characterise this \emph{orbit drift} for general particle shapes $\lambda$ by computing the Floquet exponents of the limiting orbits.

\begin{figure}
\includegraphics{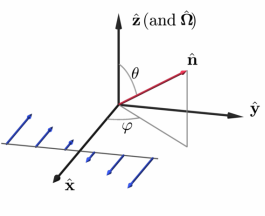}
\caption{\label{fig:coordinates} Simple shear flow and coordinate system.}
\end{figure}

\begin{figure}
\includegraphics{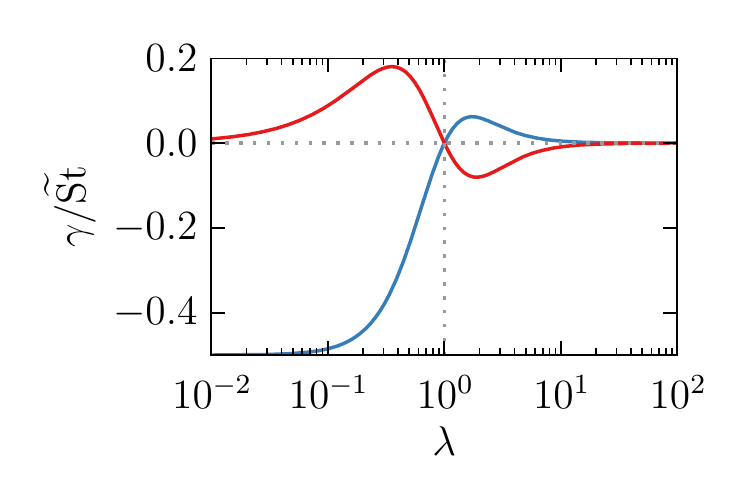}
\caption{\label{fig:lyapunovs} Floquet exponents $\gamma$ (divided by \steff) as functions of aspect ratio $\lambda$. Polar orbit, $\gamma_0$, in blue, equatorial orbit, $\gamma_{\pi/2}$, in red.}
\end{figure}

The modified vector field on the right hand side of \Eqnref{nequation} allows for two limiting orbits. First the equatorial orbit, where the $\ve n$-vector moves in the plane perpendicular to vorticity (usually referred to as \emph{tumbling}), and second the polar orbit, where the $\ve n$-vector is fixed parallel to the vorticity (referred to as \emph{log rolling}). We introduce a spherical coordinate system where $\theta$ is the polar angle, measured from the vorticity vector, and $\varphi$ is the azimuthal angle, measured from the flow direction, see Fig.~\ref{fig:coordinates}. Then the two limiting orbits correspond to $\theta=0$ and $\theta = \pi/2$. In spherical coordinates, \Eqnref{nequation} takes the following
form for a simple shear flow:
\begin{align*}
	\dot{\varphi} &= f(\theta, \varphi) = \frac{1}{2} (-1+\Lambda  \cos 2 \varphi )  \\
	&\quad + \frac{\steff \Lambda}{8} 
	 \left(2 \Lambda  \sin ^2\theta  \sin 4 \varphi +\sin 2 \varphi  ((\Lambda +1) \cos 2
   \theta +\Lambda -3)\right)\,,\\
	\dot{\theta} &= g(\theta, \varphi) = \Lambda  \sin \theta  \cos \theta  \sin \varphi  \cos \varphi   \\ 
	&\quad +  \frac{\steff \Lambda}{2} \sin \theta  \cos \theta  \cos ^2(\varphi ) \left(4 \Lambda  \sin ^2\theta  \sin
   ^2\varphi -\Lambda +1\right)\,.
\end{align*}
These equations show that inertia induces a coupling between nutation and precession, in contrast with the  inertia-free case
where the $\varphi$-dynamics is independent of $\theta$.
In the following we denote the limiting periodic orbits $\varphi_0(t)$ and $\theta_0=0$ or $\pi/2$. The solutions $\varphi_0(t)$ are in both cases periodic with the Jeffery period $T=4\pi/\sqrt{1-\Lambda^2}$. If we introduce arbitrarily small perturbations $\theta(t)=\theta_0+\delta \theta(t)$ and $\varphi(t)=\varphi_0(t)+\delta \varphi(t)$, then the perturbation $\delta \theta(t)$ evolves according to
\begin{align*}
	\frac{\rd}{\rd t} \delta \theta &=  \delta \theta \frac{\partial g}{\partial \theta}\evalat{(\varphi_0(t), \theta_0)} \!\!\!\!\!+\, \delta \varphi \frac{\partial g}{\partial \varphi}\evalat{(\varphi_0(t), \theta_0)} \,.
\end{align*}
We note that ${\partial g}/{\partial \varphi} = 0$ when evaluated at $\theta=0$ or $\theta=\pi/2$. This simplification yields
\begin{align}
	\frac{\rd}{\rd t}\ln \delta \theta &= \frac{\partial g}{\partial \theta}\evalat{(\varphi_0(t), \theta_0)}\eqnlab{lyap1}\,.
\end{align}
The Floquet exponent is obtained by time-averaging \Eqnref{lyap1} along the unperturbed periodic orbit $(\varphi_0, \theta_0)$:
\begin{align}
	\gamma_\theta &= \frac{1}{T}\int_0^{T}\rd t \frac{\partial g}{\partial \theta}\evalat{(\varphi_0(t), \theta_0)} = -\frac{1}{T}\int_0^{2\pi}\frac{\rd \varphi}{\dot \varphi} \frac{\partial g}{\partial \theta}\evalat{(\varphi, \theta_0)}\,.\eqnlab{floquet}
\end{align}
The last equality follows from the fact that in one period, $\varphi$ travels monotonically through $\varphi=0$ to $-2\pi$. Upon evaluating \Eqnref{floquet} we obtain for the polar, log rolling orbit
\begin{align*}
	\gamma_0 &= \frac{\steff}{4}\Lambda(1-\Lambda) + \mathcal O(\st^2),
\end{align*}
and for the equatorial, tumbling orbit
\begin{align*}
	\gamma_{\pi/2} &= \frac{\steff}{4}(1-\Lambda)(\sqrt{1-\Lambda^2}-\Lambda-1)+ \mathcal O(\st^2).
\end{align*}
A positive Floquet exponent $\gamma$ indicates an unstable orbit, and a negative an attracting, stable orbit. The exponents (divided by $\steff$) are shown as a function of particle shape $\lambda$ in Fig.~\ref{fig:lyapunovs}. We observe that the stable limiting orbit for flat, disk-shaped particles is the polar, log rolling orbit, while rod-shaped particles are attracted to the equatorial, tumbling orbit. This is consistent with the intuition that the most stable rotation is that with the mass distributed far from the axis of rotation. 
Our result explains the orbit drift observed in \cite{Lundell2010} at small values of \st. In Ref.~\cite{Sub06} this drift (at
small Stokes number and vanishing particle Reynolds number) was explained for nearly spherical particles. Here we generalise the result to arbitrary axisymmetric particles.

\section{Competition between noise and inertia}
The indeterminacy of the Jeffery orbits poses a problem for predictions involving the orientational distribution of particles. Jeffery \cite{Jeff22} hypothesised that orbits corresponding to minimal energy dissipation should be preferred. Another approach, pursued by a number of authors \cite{Sch55, Hin72, Bre74}, shows that thermal noise, after a long time, removes the dependence on initial conditions and produces a stationary distribution of the $\ve n$ vector. In particular, in \cite{Hin72} it is shown that the orientational distribution converges to a function independent of noise strength, as the noise strength is reduced. On the contrary, in the preceding sections we have shown that in the absence of noise a weakly inertial particle will rotate into a specific stable orbit. In this section we consider what effect weak particle inertia have on the orientational distribution.

The time scale for inertia to have an effect is equal to the inverse Floquet exponent, $s /\gamma \sim 1/\steff$. The corresponding time scale for diffusion is $s \tau_D \sim \pe$, where the P\'eclet number $\pe=s/D$ measures the noise strength, or diffusion constant $D=k_B T/\widetilde{Y}^C$, relative to the shear strength \cite{Kim91}. Thus, when $\tau_D\gamma \sim \pe \steff \ll 1$ we expect noise to determine the stationary distribution, as described by previous authors \cite{Bre74}. On the other hand, when $\pe\steff\gg1$ we expect a distribution dominated by the inertial effects, with a peak around the stable orbit described in the previous section. In order to quantify these expectations we need to compute moments of the stationary distribution for various $\pe$, $\steff$ and $\lambda$.

The Fokker-Planck equation describing the evolution of the orientational probability density $P(\ve n, t)$ is
\begin{align}
	&\partial_t P(\ve n, t) = -\partial_{\ve n}\left[(F_0(\ve n) + \st F_1(\ve n))P\right] + \pe^{-1}\partial^2_{\ve n}P \equiv \hat J P. \eqnlab{fpe} 
\end{align}
Here $\partial_{\ve n}$ is the gradient on the sphere defined as $\partial_{\ve n} \equiv (\maid - \venn)\nabla$. The boundary condition of \Eqnref{fpe} is that the probability density $P$ must integrate to unity over the sphere. The Fokker-Planck equation can be solved numerically by expansion in spherical harmonics, the details of this procedure are described in \ref{app:fpe}.

For a discussion of the solutions of \Eqnref{fpe} we compute the stationary average $\langle \sin^2\theta\rangle$, where $\theta$ is the polar angle from the vorticity vector of the simple shear flow (see Fig.~\ref{fig:coordinates}). Recall that weakly inertial rod-shaped particles exhibit a stable orbit at $\theta = \pi/2$, and disk-shaped particles at $\theta=0$. Moreover, values of $\langle \sin^2\theta\rangle$ have been tabulated (Table~6c in Ref.~\cite{Bre74}) for different values of $\pe$ and $\lambda$. This allows us to validate our method at $\st=0$.

\begin{figure}
\includegraphics{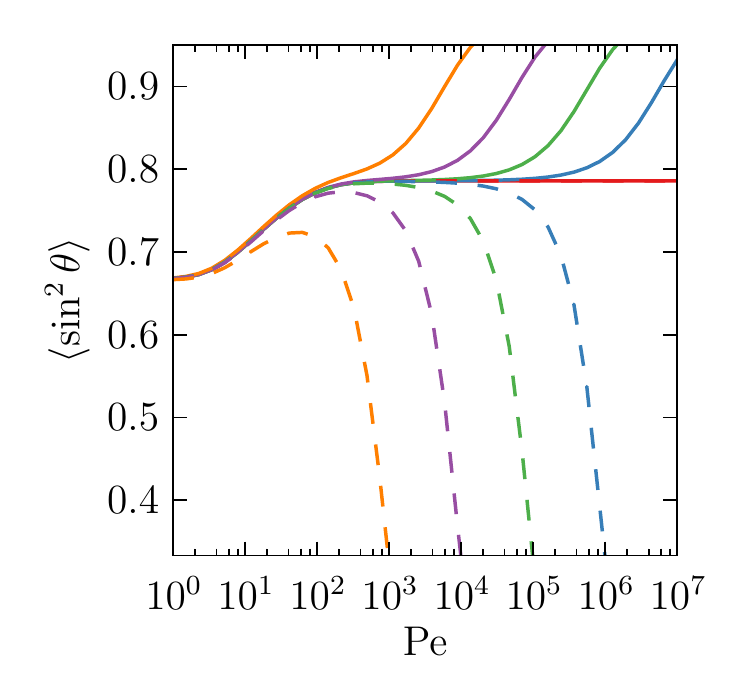}
\caption{\label{fig:sintheta1} Stationary orientational average as function of $\pe$ for different $\steff$ and $\lambda$. Solid lines are a prolate particle $\lambda=5$, dashed lines oblate $\lambda=1/5$. The red solid line is $\steff=0$ result, then blue $\steff = 10^{-4}$, green $\steff = 10^{-3}$, purple $\steff = 10^{-2}$ and orange $\steff = 10^{-1}$. }
\end{figure}

In Fig.~\ref{fig:sintheta1} we show the angular average  $\langle \sin^2\theta\rangle$ as a function of $\pe$ for different values of $\st$, and for two particle shapes: prolate $\lambda=5$ and oblate $\lambda=1/5$. We note the following features. At strong noise, that is small $\pe$, the moment is independent of $\st$ and approaches the uniformly random result $\langle \sin^2\theta\rangle=2/3$. Next we see that the curve for $\st=0$ (red curve) levels out and converges at a finite value of $\pe$, as predicted in Ref.~\cite{Hin72}. Finally we see that for finite $\st$, the curves coincide with the $\st=0$ curve up to $\pe\st\approx1$, but then deviates and instead converges towards $\langle\sin^2\theta\rangle=1$ (prolate particle), or $\langle\sin^2\theta\rangle=0$ (oblate particle) when $\pe \st \gg 1$.

\begin{figure}
\includegraphics{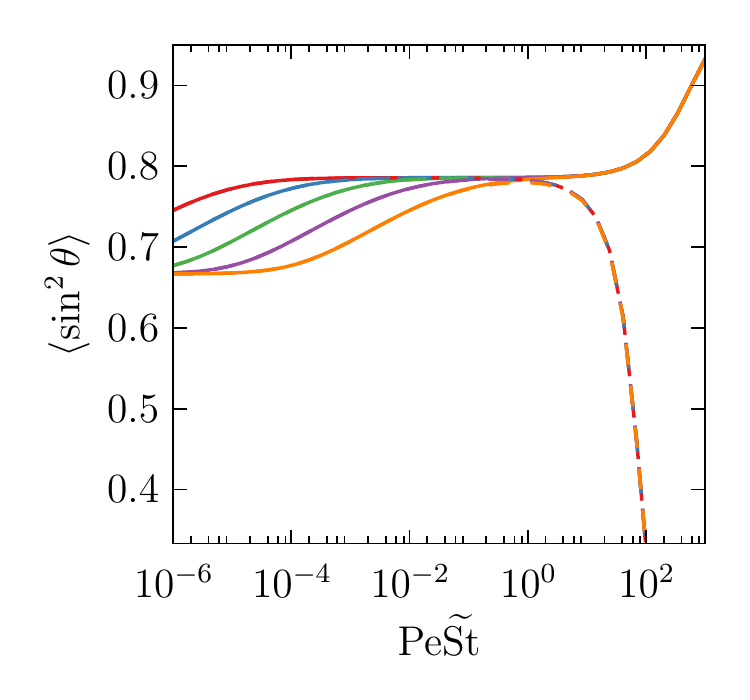}
 \caption{\label{fig:sintheta2} Stationary orientational average as function of $\pe\steff$ for different values of $k=\steff/\pe$. Solid lines are a prolate particle $\lambda=5$, dashed lines oblate $\lambda=1/5$. Then red $k = 10^{-9}$, blue $k = 10^{-8}$, green $k = 10^{-7}$, purple $k = 10^{-6}$ and orange $k = 10^{-5}$.}
 \end{figure}

 The limit of $\st\to0$ is singular: the stationary distribution at low noise and weak inertia differs substantially from the one predicted by a $\st=0$ calculation, although the time to arrive at the steady state becomes longer for weaker inertia. The most important observation is that the stationary average at low noise strength but without inertia is the same for prolate and oblate particles, whereas the inertial correction has opposite directions for the two different shapes. We emphasise this observation in Fig.~\ref{fig:sintheta2}, where we show the stationary average as function of $\pe \steff$, for different values of $\steff/\pe$. It turns out that when $\pe \steff$ is large enough, then the average does not depend on $\steff$ and $\pe$ separately, but only in the combination $\pe \steff$. The stationary average is determined by the ratio of relaxation times $\tau_D\gamma \sim\pe \steff$, and by the aspect ratio $\lambda$. For small values of $\pe \steff$ the stationary distribution is instead determined by $\pe$, as shown in Fig.~\ref{fig:sintheta1}.

\section{Conclusions}
\label{sec:conc}

The orientational dynamics of weakly inertial axisymmetric particles in a steady flow has been investigated. 
We have derived an approximate equation of motion for the unit
axial vector along the particle symmetry axis, assuming that the Stokes number is small. 
The fact that we have obtained this equation of motion in explicit form (valid for any axisymmetric and weakly inertial particle in any steady linear viscous flow) has
made it possible to address two important questions, both pertaining to the case of a simple shear flow. 

First, we have shown that axisymmetric particles drift towards either the log-rolling (disk-shaped particles) or the tumbling orbit (rod-like particles). We have quantified 
this result for any particle aspect ratio by calculating the Floquet exponents of the periodic orbits to leading order in $\st$. 
Our result generalises a result for nearly spherical particles in Ref.~\cite{Sub06} to arbitrary axisymmetric particles.

Second, the approximate equation of motion derived in Sec.~2 has allowed us to formulate the Fokker-Planck equation describing the orientational distribution of axisymmetric, weakly inertial particles in a shear flow when rotational noise is present. Numerical solutions of the Fokker-Planck equation show that stationary orientational distributions are close to the inertia-free case when $\pe\steff\ll1$, whereas they are determined by inertial effects, though small, when $\pe\gg1/\steff\gg1$.

An important open question is to describe the effect of fluid inertia (at small but finite particle Reynolds number) upon the orientational motion for particles with arbitrary
aspect ratios, in competition with the effects of particle inertia and rotational noise. 

The calculations summarised in this article are restricted to axisymmetric particles. But our approach is readily generalised to asymmetric (triaxial) particles. 

\section*{Acknowledgements}
Financial support by Vetenskapsr\aa{}det, by
the G\"oran Gustafsson Foundation for Research in Natural Sciences and Medicine, and by the EU
COST Action MP0806 on \lq Particles in Turbulence' are
gratefully acknowledged.

\clearpage

\appendix
\section{Solution of the Fokker-Planck equation}\label{app:fpe}
We start from the Fokker-Planck equation governing the evolution of the probability density $P(\ve n, t)$ of finding a particle with orientation vector $\ve n$ at time $t$:
\begin{align}
	\partial_t P(\ve n, t) &= -\partial_{\ve n}\left[\dot{\ve n}P\right] + \pe^{-1}\partial^2_{\ve n}P \equiv \hat J P \eqnlab{fpeapp} \\
	\intertext{with the normalisation condition}
	\int_{S_2} \!\!\!\rd \ve n\,P(\ve n, t) &= 1.
\end{align}
The differential operator $\partial_{\ve n}$ is the gradient on the sphere, defined by taking the usual gradient $\nabla$ in $\mathbb R^3$ projected onto the unit sphere, $\partial_{\ve n} \equiv (\maid - \venn)\nabla$. The drift term $\dot{\ve n}$ may be any polynomial function of $\ve n$, for example the equation of motion \Eqnref{nequation}.

We approximate the solution of \Eqnref{fpeapp} by an expansion in spherical harmonics \cite{Sch55}, the eigenfunctions of the quantum-mechanical angular-momentum operators, which form a complete basis on $S_2$. We use bra-ket notation,
\begin{align}
	P(\ve n, t) &= \langle \ve n| P(t)\rangle =  \sum_{l=0}^{\infty}\sum_{m=-l}^l c_l^m(t) \langle \ve n\Ylm{l}{m}\eqnlab{basisexpansion}
\intertext{where}
	 \langle\ve n|l,m\rangle &= Y_l^m(\ve n) = (-1)^m\sqrt{\frac{2l+1}{4\pi}\frac{(l-m)!}{(l+m)!}}P_l^m(\cos\theta)e^{im\varphi}.
\end{align}
We use the standard spherical harmonics defined in for example \cite{Arf70} (p. 571). The functions $P_l^m$ are the associated Legendre polynomials. We call the time-dependent coefficients for each basis function $c_l^m(t)$, and the fact that $P$ is real-valued constrains the coefficients: 
\begin{align}
	c_l^{-m} = (-1)^m\overline{c_l^m},\eqnlab{clmreal}
\end{align}
where the bar denotes complex conjugation. Then \Eqnref{fpeapp} for the time evolution of the state $|P(t)\rangle$ reads
\begin{align*}
	\frac{\rd}{\rd t} |P(t)\rangle &= \hat J |P(t)\rangle\,.
\end{align*}
Inserting the expansion yields
\begin{align}
	\sum_{l=0}^{\infty}\sum_{m=-l}^l \frac{\rd}{\rd t} c_l^m(t) \Ylm{l}{m} &= \sum_{l=0}^{\infty}\sum_{m=-l}^l c_l^m(t) \hat J\Ylm{l}{m}.
	\intertext{Multiplying with $\langle p,q|$, and using the orthogonality of the spherical harmonics we arrive at a system of coupled ordinary differential equations for the coefficients $c_l^m(t)$}
	\dot c_p^q(t)  &= \sum_{l=0}^{\infty}\sum_{m=-l}^l c_l^m(t) \langle p,q| \hat J\Ylm{l}{m} \eqnlab{clmeq1},\\
	c_0^0 &= \frac{1}{\sqrt{4\pi}}\eqnlab{clmeq1norm}.
\end{align}
In order to solve \Eqnref{clmeq1} it remains to compute the matrix elements of the operator $\hat J$. This is achieved by expressing $\hat J$ as a combination of the angular momentum operators. How the angular momentum operators act on the spherical harmonics is well known, see for example Ref.~\cite{Arf70}.

The angular momentum operator $\hat{\ve L}$ is given by
\begin{align*}
	\hat{\ve L} &= -i \hat{\ve n}\cross\nabla.
\intertext{In terms of $\hat{\ve L}$ we have}
	\partial_{\ve n}^2 &= (\maid-\hat{\ve n}\hat{\ve n}\transpose)\nabla \cdot (\maid-\hat{\ve n}\hat{\ve n}\transpose)\nabla = - \hat{\ve L}^2,
\intertext{and}
	\partial_{\ve n}&= (\maid-\hat{\ve n}\hat{\ve n}\transpose)\nabla =  -i\hat{\ve n} \cross \hat{\ve L}.
\end{align*}
The states $\Ylm{l}{m}$ are the eigenfunctions of $\hat{\ve L}^2$ and $\hat{L}_3$ with eigenvalues $l(l+1)$ and $m$. Now, to evaluate the drift term $\partial_{\ve n}\dot{\ve n}$ we need to know how $\hat{L}_1$, $\hat{L}_2$, and $\hat{\ve n}$ act on $\Ylm{l}{m}$. The first two are known through the use of ladder operators, defined by
\begin{align*}
	\hat{L}_{\pm} = \hat{L}_1 \pm i\hat{L}_2 \implies \hat{L}_1 = \frac{1}{2}(\hat{L}_+ + \hat{L}_-),\quad \hat{L}_2 = -\frac{i}{2}(\hat{L}_+ - \hat{L}_-)
\end{align*}
and their effect is
\begin{align*}
	\hat{L}_\pm \Ylm{l}{m} &= \sqrt{(l\mp m)(1+l\pm m)}\,\Ylm{l}{m\pm1}.
\end{align*}
Next, the drift term we consider is a polynomial in $\hat{n}_1, \hat{n}_2$ and $\hat{n}_3$ (the components of $\hat{\ve n}$), we therefore need to evaluate the action of a monomial $\hat{n}_1^\alpha \hat{n}_2^\beta \hat{n}_3^\gamma$ on $\Ylm{l}{m}$. Any such monomial of order $k=\alpha+\beta+\gamma$ may be written as a linear combination of spherical tensor operators of up to order $k$:
\begin{align*}
	\hat{n}_1^\alpha \hat{n}_2^\beta \hat{n}_3^\gamma &= \sum_{l=0}^{k}\sum_{m=-l}^l a_l^m(\alpha,\beta,\gamma)\hat{Y}_l^m.
\end{align*}
The action of $\hat{Y}_p^q$ is computed with the Clebsch-Gordan coefficients (see for example Ref.~\cite{Sak94}, p.216)
\begin{align*}
	\hat{Y}_p^q \Ylm{l}{m} &= \sum_{\Delta l=-p}^{p}K(p,q,l,m,l+\Delta l, m+q)\Ylm{l+\Delta l}{m+q},\\
\end{align*}
where
\begin{align*}
	K(l_1, m_1, l_2, m_2, l, m) &= \sqrt{\frac{(2l_1+1)(2l_2+1)}{4\pi (2l+1)}} \\
	&\quad \times \langle l_1 l_2; 0 0|l_1l_2;l0\rangle
	\langle l_1 l_2; m_1 m_2|l_1l_2;lm\rangle\,.
\end{align*}
The Clebsch-Gordan coefficients are denoted by $\langle l_1 l_2; m_1 m_2|l_1l_2;lm\rangle$ (see e.g Eq. (3.7.44) of Ref.~\cite{Sak94}), and they are available in Mathematica by the function \texttt{ClebschGordan[{l1,m1},{l2,m2},{l,m}]}. Finally, we order the operators, so that as many terms as possible cancel before we actually begin evaluating the operators.
In particular we want to reorder $\hat{n}_i$ against $\hat{L}_i$, and we make use of the commutator
\begin{align*}
	[\hat{n}_p, \hat{L}_q] &=\sum_{j=1}^3 i\varepsilon_{pqj}\hat{n}_j,
\end{align*}
where $\varepsilon_{ijk}$ is the Levi-Civita tensor.

Let us consider an example. In the special case of $\hat J = \pe^{-1} \partial_{\ve n}^2$ we see that $\hat J=-\pe^{-1} \hat{\ve L}^2$. It follows:
\begin{align*}
	\hat J \Ylm{l}{m}&=-\pe^{-1}\hat{\ve L}^2 \Ylm{l}{m} = -l(l+1)\pe^{-1}\Ylm{l}{m}.
\end{align*}
Knowing how $\hat J$ acts on $\Ylm{l}{m}$ fully specifies \Eqnref{clmeq1} which becomes
\begin{align*}
	\dot c_p^q(t)  &= -p(p+1)\pe^{-1}c_p^q(t).
\end{align*}
This means that the diffusion operator exponentially suppresses all modes with $p>0$, and the lowest mode $p=0$ is determined by the normalisation condition \Eqnref{clmeq1norm}. This is the solution of the diffusion equation on the sphere, with the uniform distribution as steady state.

Now take the for the drift term the standard Jeffery equation $\dot {\ve n} = \ma O \ve n + \Lambda (\ma S \ve n - \venn \ma S \ve n) = \ve F_0(\ve n)$, and call this operator $\hat J = -\partial_{\ve n}\ve F_0(\ve n) + \pe^{-1}\partial^2_{\ve n}$. We find
\begin{align*}
	\hat J &= 
	i \sqrt{\frac{\pi }{30}} \Lambda  \hat{L}_-\hat{Y}_{2}^{-1}+i \sqrt{\frac{\pi }{30}} \hat{L}_-\hat{Y}_{2}^{1}-i \sqrt{\frac{\pi }{30}} 
   \Lambda  \hat{L}_+\hat{Y}_{2}^{1} \\ 
   &\quad-  i \sqrt{\frac{\pi }{30}} \hat{L}_+\hat{Y}_{2}^{-1} 
   -i \sqrt{\frac{2 \pi }{15}} \Lambda 
   \hat{L}_3\hat{Y}_{2}^{-2}-i \sqrt{\frac{2 \pi }{15}} \Lambda  \hat{L}_3\hat{Y}_{2}^{2} \\ 
   &\quad
   +\frac{2}{3} i \sqrt{\pi } 
   \hat{L}_3\hat{Y}_{0}^{0}-\frac{2}{3} i \sqrt{\frac{\pi }{5}} \hat{L}_3\hat{Y}_{2}^{0}-\pe^{-1} \hat{\ve L}^2.
\end{align*}
The fact that the operator $\hat J$ is real implies that the matrix elements must have the symmetry
\begin{align*}
	\langle p, q|\hat J |l,m\rangle = (-1)^q \overline{\langle p, -q|\hat J |l,m\rangle}.
\end{align*}
We can understand the reason for this because the system of differential equations in \Eqnref{clmeq1} needs to preserve the condition that $P$ is real, as stated in \Eqnref{clmreal}. It implies we have only to compute half of the matrix elements.

Upon including the correction \Eqnref{nequation} due to weak particle inertia, the expression for the operator $\hat{J}$ becomes too lengthy to include here. However, we have implemented the identities and rules described in this appendix in a Mathematica notebook \cite{githubrepo}. Given an operator $\hat J$ expressed in terms of $\hat{\ve n}$ and $\hat{\ve L}$, it converts the expression into sums of angular momentum operators as exemplified above. We also provide an additional Mathematica notebook that, given the matrix elements, assembles a sparse matrix $\ma J$, up to a desired order $l_{\textrm{max}}$. This matrix can then be used to solve the truncated version of \Eqnref{clmeq1}:
\begin{align*}
	\dot {\ve c} &= \ma J \ve c,
\end{align*}
where $\ve c$ is a vector with elements $c_l^m$.
Alternatively, the matrix can be solved for the stationary solution of $\ve c$. 
All results shown in this paper were computed using $l_{\mathrm{max}}=400$, which leads to very good convergence in all cases shown. Only when solutions approach delta peaks, as for example the limiting stable orbits in the large $\pe\st$ case, the expansion procedure converges very slowly.
\end{document}